\documentclass[
              twocolumn,
               noshowpacs,          
               nopreprintnumbers,     
               aps,                 
               prd,                 
               superscriptaddress,  
               nofootinbib,         
               tightenlines,        
               floats,floatfix      
               ]{revtex4}

\usepackage{amsmath,amssymb}
\usepackage{feynmp}
\usepackage{float}
\usepackage{setspace}
\usepackage{latexsym}
\usepackage{graphicx,color}
\usepackage{epstopdf}
\usepackage{graphicx}
\usepackage[font=small,skip=0pt]{caption}
\usepackage{makeidx} 
\usepackage{lmodern}
\usepackage{enumitem}

\usepackage[T1]{fontenc}
\usepackage[utf8]{inputenc}
\usepackage{color}
\usepackage{booktabs}
\usepackage{units}
\usepackage{amsmath}
\usepackage{amssymb}
\usepackage{esint}

\def\be{\begin{equation}}
\def\ee{\end{equation}}
\def\ba{\begin{eqnarray}}
\def\ea{\end{eqnarray}}

\def\bdm{\begin{displaymath}}
\def\edm{\end{displaymath}}

\def\bq{\begin{quote}}
\def\eq{\end{quote}}

 at 10truept
\newcommand{\del}{\partial}

\newcommand{\beq}{\begin{equation}}
\newcommand{\eeq}{\end{equation}}
\newcommand{\bea}{\begin{eqnarray}}
\newcommand{\eea}{\end{eqnarray}}
\newcommand{\beqa}{\begin{eqnarray}}
\newcommand{\eeqa}{\end{eqnarray}}

\newcommand{\AP}[1]{{ #1}} 

\def\ltap{\ \raise.3ex\hbox{$<$\kern-.75em\lower1ex\hbox{$\sim$}}\ }
\def\gtap{\ \raise.3ex\hbox{$>$\kern-.75em\lower1ex\hbox{$\sim$}}\ }
\def\gl{\ \raise.5ex\hbox{$>$}\kern-.8em\lower.5ex\hbox{$<$}\ }
\def\roughly#1{\raise.3ex\hbox{$#1$\kern-.75em\lower1ex\hbox{$\sim$}}}

\begin{document}

\title{Unitarity and the Vainshtein Mechanism}

\author{Nemanja Kaloper} \email{kaloper@physics.ucdavis.edu}
\affiliation{Department of Physics, University of California, Davis, CA95616, USA} 
\author{Antonio Padilla}\email{antonio.padilla@nottingham.ac.uk} \author{Paul M. Saffin} \email{paul.saffin@nottingham.ac.uk}
\author{David Stefanyszyn} \email{ppxds1@nottingham.ac.uk}
\affiliation{School of Physics and Astronomy, 
University of Nottingham, Nottingham NG7 2RD, UK} 

~

\vskip.2cm

~
~

\date{\today}

\begin{abstract}
We investigate low energy limits of massive gauge theories that feature the Vainshtein mechanism, focussing on the effects of the UV modes that are integrated out. It turns out that the Goldstone sectors are significantly influenced by the effects from such modes relative to the effective field theories where the irrelevant operators induced by heavy modes are simply cast aside. The effects of the consistently retained higher order corrections affect the strong coupling and show that the nature of the UV completion influences the low energy theory significantly. This casts doubts on the 
naively estimated environmental strong coupling scale, and on the effectiveness of the environmental enhancement of screening. The environmental effects by themselves might not suffice to cure the bad behavior of the theory beyond the vacuum cutoff.
\end{abstract}
\maketitle

\section{Introduction}

The expansion of the universe appears to be accelerating. This phenomenon could be accounted for by a small amount of dark energy, i.e. a tiny cosmological constant, or alternatively by a modification of General Relativity (GR) (for recent reviews see \cite{rev1,rev2}). In the latter case, some kind of `screening' mechanism is needed to retain the experimental successes of GR in the tested limits. 
Many key features of modifications of gravity that appear in the literature are encapsulated in simpler models involving a single long range scalar field, $\varphi$, with nonminimal couplings. In this context,  screening mechanisms  are simply understood by considering the dynamics of scalar fluctuations, $\delta \varphi$,  about some environmentally dependent background, $\bar \varphi$. At the quadratic order these are typically described by a Lagrangian of the form
\be
\delta {\cal L}=-\frac12 Z^{\mu\nu}[\bar\varphi] \del_\mu \delta \varphi \del_\nu \delta \varphi-\frac12 M^2[\bar\varphi] \delta \varphi^2+\frac{1}{M_{pl}} \delta \varphi \delta T \, ,
\ee
where $Z^{\mu\nu}$ is an environmentally dependent $Z$ factor that alters the effective coupling, and $M$ is an environmentally dependent mass. Generically, when $Z^{\mu\nu} \sim {\cal O}(1)$, the scalar couples to the fluctuation in the trace of the energy-momentum tensor, $\delta T$, with gravitational strength.  The chameleon mechanism \cite{cham} works by rendering the mass $M$ large in the Earth's environment, such that the range of the scalar force is less than the millimetre scale probed by table top tests of gravity. In contrast, the Vainshtein mechanism \cite{vain,dyson} exploits an environmental increase in the $Z$ factor, $Z^{\mu\nu} \gg 1$ so that the effective coupling to matter is much weaker than gravity, $\frac{1}{M_{pl}\sqrt{Z}} \ll \frac{1}{M_{pl}}$.

In this paper, we will study  Vainshtein screening  in  weakly coupled non-Abelian massive gauge theories. These theories have a strong coupling cutoff at a scale $\sim m_A/g$, but can be UV completed by the Higgs mechanism. Our main interest will be to see the IR effects of the extra UV modes in the
completion, namely the radial mode. For simplicity, our strategy to analyze the dynamics will be to focus on the helicity-0 sector of the gauge theory.
By resorting to the Goldstone equivalence theorem, the dynamics of the helicity-0 sector can be extracted by the Stuckleberg method, and the transverse modes can be ignored. The Stuckelberg fields then come in the form of a non-linear $\sigma$-model, which for the simplest massive gauge theory is just $SU(2)$. We will view this setup as an avatar of what one may expect for massive gravity, at least in terms of the flavor of problems one may encounter. While the precise proof that this is the case is currently absent, heuristics suggests this philosophy is reasonable. Thus, in practice we will be focusing on the $\sigma$-model scalar field theories, including the $U(1)$ $\sigma$-model as the simplest case, which feature the  Vainshtein mechanism.

The environmental dependence of the $Z$ factor is achieved in the full theory via derivative self-interactions of the field.  These interactions manifest themselves through the breakdown of classical perturbation theory: when the non-linearities kick in and become important, the $Z$ factor is enhanced.  In a static, spherically symmetric scenario, for this to occur at a macroscopic scale beyond the Schwarzschild radius of the source, the scale  suppressing the self-interactions should be well below the Planck scale $\Lambda \ll M_{pl}$. Quantum mechanically this suggests that the interaction generically becomes strong at  the same scale $\Lambda$, signaling the breakdown of perturbative unitarity and the limit of the (truncated) effective field theory (EFT) description. In standard QFT, the appearance of such behavior is taken to point to the necessity of a Wilsonian completion at the scale $\Lambda$, which extends the low energy theory beyond the cut-off.  This generically requires new degrees of freedom near the cut-off that help to soften the interactions while maintaining unitarity (for a textbook discussion see e.g. \cite{georgi}). A simple example of this is the Higgs, which preserves unitarity in WW scattering above the strong coupling scale of the massive non-Abelian gauge theory. For an illustration of the equivalent scales at which such a UV completion should be relevant in a modified gravity context, note that for Lorentz invariant massive gravity the EFT cut-off is at the scale $\Lambda_3\sim (m^2 M_{pl})^{1/3}$ \cite{nima}, which for a Hubble mass graviton yields the macroscopic scale $\Lambda_3 \sim (1000\text{km})^{-1}$.

Of course, having a gravity theory that is only valid down to $1000$km or so is unsatisfactory to say the least.  It has been suggested that the truncated low energy effective description can be maintained beyond the cut-off, $\Lambda$,  in the presence of a non-trivial environment when Vainshtein screening is active \cite{nic}. This is precisely because of the enhanced $Z$ factor, which serves to weaken the effective coupling for fluctuations about the classical environment.  The interactions describing these fluctuations now become strong at a much higher scale, $\Lambda_\text{env} \gg \Lambda$.  For massive gravity this improved things a little (although not enough), with the environmental strong coupling scale in the field of the Earth rising as high as $\Lambda_\text{env} \sim 1/\text{km}$ \cite{strong}. For DGP gravity the environmental cut-off seems better, near  the centimeter scale \cite{nic}.

Implicit in all of these discussions of environmental strong coupling scales is the assumption that whatever completes the theory beyond the scale $\Lambda$ remains unimportant up to the new cut-off $\Lambda_\text{env}$.  For example, if we imagine that a particular theory can be completed by integrating in a heavy particle with mass $m \lesssim \Lambda$, then we are assuming that for some reason this particle remains more or less inert up to the new strong coupling scale. This, in fact, seems a bit too optimistic. Indeed, if we properly integrate out a heavy field, at scales below its mass we would expect to find many {\it  additional} higher dimensional operators suppressed by the mass (for recent discussions, see e.g. \cite{cliff}). Classically, once the first non-linearity has become important as required by the Vainshtein mechanism, it seems likely that these additional terms will also begin to kick in at some macroscopic scale at least as big as $1/m$, spoiling the classical solution to the truncated effective theory beyond that scale. If so, the classical solution may not be reliable, and thus $\Lambda_\text{env}$ calculated in the quadratic truncation may cease to be physically sensible, having been computed on a wrong background. 

Our purpose here is to explore this in more detail, and investigate the signatures of a UV completion on the classical solution with Vainshtein screening.  For the examples that we study, we will find that the truncated low energy EFT cannot be trusted much beyond the scale at which Vainshtein screening occurs. Extrapolating this result one might worry that generic theories with Vainshtein screening, including modified gravity, cannot really be taken seriously beyond the vacuum strong coupling scale 
until the full effect of their (partial) UV completion is known.  This in fact is not counter to the philosophy of \cite{nic}, where the argument was not that DGP was well behaved much above the vacuum strong coupling scale. On the contrary, the point was that the effects indicating the breakdown of unitarity of the low energy theory would merely remain well hidden from the limited direct tests of GR at short distances.

At first glance, finding a UV completion of a theory with Vainshtein screening might seem a little optimistic. Indeed, for the standard set-up in which screening occurs for static, spherically symmetric profiles around a heavy source, we typically find that fluctuations about the background exhibit superluminality \cite{uvir,deser}. This has been associated with an inability to find a standard Wilsonian completion \cite{uvir} \AP{(although see also \cite{Shore})}. Since there are no complete, compelling alternatives to the Wilsonian method to date\footnote{The so-called {\it non-Wilsonian completions} \cite{class} attempt to look to high energies in a way that seems to be different, but an argument has been offered that such non-Wilsonian completions are in fact Wilsonian in principle, but incomplete \cite{kovner}.}, we will look for explicit examples of low energy theories that feature Vainshtein screening with a standard Wilsonian UV completion that cures the aforementioned problems. In detail, for the most part we will focus on Vainshtein screening in a homogeneous environment in which the $Z$ factor grows with the energy of the background. Superluminal pathologies are avoidable with a proper UV completion, and in the examples we will consider this will be explicit. We will compare the dynamics of the truncated low energy EFT, and its UV completion, assuming the same initial data and the same energy density. For the UV completion, we will need to specify initial conditions for the new degrees of freedom that are now present, and in considering solutions we will scan over all consistent possibilities. 

Our main result is that the two descriptions begin to deviate significantly from one another whenever the energy density exceeds the scale at which Vainshtein screening kicks in. This seems generic, suggesting that classical solutions to theories with Vainshtein screening cannot be trusted much beyond the Vainshtein scale, at least not until the full effect of the (partial) UV completion is known. Special initial conditions yielding configurations that appear more stable exist, but they require tuning of parameters.
 
The rest of this paper is organized as follows: we begin in section \ref{sec:U1} with a detailed study of Vainshtein screening in a simple {\it K-essence} like model, admitting a UV completion in terms of a $U(1)$ Higgs - which is really a UV completion of a massive Abelian gauge theory (which is, clearly, not unique). This will represent the main body of the paper in which generic features are also discussed. We will show that the classical  homogeneous solution for the truncated low energy EFT cannot be trusted beyond the cut-off $\Lambda$, above which screening is meant to kick in. The low energy theory merely describes a phase - the Goldstone mode - whereas the full UV description also contains the radial mode that begins to significantly affect the dynamics for solutions with energy density ${\cal E }\gg \Lambda^4$.  In section \ref{sec:su2}, we perform a similar, although less detailed, analysis for a model of Vainshtein screening that can be completed in an $SU(2)$ $\sigma$-model,  which, as noted, comprises a UV completion of the
massive $SU(2)$ gauge theory. Our results are essentially the same as those in section \ref{sec:U1}.  In section \ref{sec:gb}, we discuss Einstein-Gauss Bonnet gravity in $D>4$ dimensions, as a stringy extension of $D$-dimensional GR. Static spherically symmetric solutions to the latter low energy theory are known to deviate from their linearized description at the Schwarzschild radius around a heavy source. This is entirely analogous to the breakdown of the linearized scalar description in modified gravity scenarios with a Vainshtein radius. We then show that the UV correction coming from Gauss-Bonnet manifests itself at a macroscopic scale just below the Schwarzschild radius. In section \ref{sec:conc} we summarize our findings, which suggest that the Vainshtein mechanism for the cases we explored is not reliable much beyond the scale of the non-linearities in the truncated low energy description.

\section{Vainshtein $\&$ Higgs: $U(1)$} \label{sec:U1}
\AP{It turns out that the important features of the non-Abelian Vainshtein screening, and its UV completion,  also appear in a much simpler Abelian scenario. With this in mind, } let us start by considering a {\it K-essence} like  theory
\be
{\cal L}=X+\epsilon \frac{X^2}{\Lambda^4}+ \frac{1}{M_{pl}}  \varphi  T,
\ee
where $X=-\frac12 (\del \varphi)^2$, $\epsilon=\pm1$, and $\Lambda \ll M_{pl}$ is some scale associated with the breakdown of perturbative unitarity for fluctuations about the vacuum.  On a non-trivial background, the fluctuations acquire a $Z$ factor
\be
Z^{\mu\nu}=\eta^{\mu\nu}\left(1+2\epsilon \frac{\bar X}{\Lambda^4}\right)-2\epsilon \frac{\del^\mu \bar \varphi \del^\nu \bar \varphi}{\Lambda^4}.
\ee
It is instructive to recall the intimate connection between the breakdown of perturbative unitarity at the scale $\Lambda$ and the breakdown of classical perturbation theory in the presence of a source \cite{nima,nonpert,predictive}. For this example, we have a canonical propagator  $\sim 1/p^2$, where $p$ is the 4-momentum, and  a 4-point self interaction of the field with a vertex that schematically scales as $p^{4}/\Lambda^{4}$.
Assuming a static and spherically symmetric profile, we now consider the classical field $\varphi^c$ outside a heavy  source of mass $M$.  Diagrammatically, we can describe this as follows,  
\begin{equation}  \label{pot}
\unitlength=1mm
\begin{fmffile}{phipic}
\parbox{15mm}{
\begin{fmfgraph*}(15,15)
\fmfleft{pl}
\fmfright{pr}
\fmf{dashes}{pl,pr}
\fmfblob{0.2w}{pl}
\end{fmfgraph*} } = \quad
%
%
\parbox{15mm}{
\begin{fmfgraph*}(15,15)
\fmfleft{pl}
\fmfright{pr}
\fmf{dashes}{pl,pr}
\fmfdot{pl}
\end{fmfgraph*} } + \quad
%
%
%
%
\parbox{15mm}{
\begin{fmfgraph*}(15,15)
\fmfleft{p1,p2,p3}
\fmfright{pr}
\fmf{dashes}{p1,pl}
\fmf{dashes}{p2,pl}
\fmf{dashes}{p3,pl}
\fmf{dashes}{pl,pr}
\fmfdot{p1}
\fmfdot{p2}
\fmfdot{p3}
\end{fmfgraph*} } +  \ldots
\end{fmffile}
\end{equation}
The first diagram contributes the usual profile for a massless scalar $\varphi_{lin}^c \sim \frac{M}{M_{pl}}\frac{1}{r}$. The next diagram yields a correction that is easily computed by dimensional analysis, $\Delta \varphi^c \sim \left( \frac{M}{M_{pl}}\right)^{3}\frac{1}{(\Lambda r)^{4}} \frac{1}{r}$.  Note that the linearized approximation becomes a poor one when $\Delta \varphi^c/\varphi^c_{lin} \sim 1$, or equivalently when $r \sim r_V \sim \frac{1}{\Lambda}\left(\frac{M}{M_{pl}}\right)^{\frac12}$. Below the Vainshtein radius $r_V$, one should sum all the diagrams in Eq. (\ref{pot}), which is of course, equivalent to solving the classical field equations exactly in the presence of the source. In other words, solve
\be
\varphi'-\epsilon \frac{\varphi'^3}{\Lambda^4} =\frac{M}{M_{pl}} \frac{1}{4\pi r^2}.
\ee
It is well known that the solution that asymptotes to $\varphi_{lin}=-\frac{M}{M_{pl}} \frac{1}{4\pi r}$ can only be continuously extended to much shorter distances when $\epsilon=-1$. Then the solution is screened for $r \ll r_V  \sim  \frac{1}{\Lambda}\left(\frac{M}{M_{pl}}\right)^{\frac12}$, scaling as $\varphi_{nonlin} \sim \left(\frac{M}{M_{pl}} \Lambda^4  r\right)^{1/3}$, and giving a $Z$ factor that goes as $Z \sim \left(\frac{r_V}{r}\right)^{4/3} \gg 1$. In contrast, when $\epsilon=+1$, the solution runs into a branch-cut singularity close to the  same scale  $r_V$. This behavior appears to be generic in models with Vainshtein screening on static backgrounds. The sign of the higher dimensional operator that kicks in at some macroscopic scale around a heavy source is critical.  For one particular sign we can extend the static, spherically symmetric solution to shorter distances, while for the other sign the solution is cut-off at the Vainshtein radius by a branch cut (see, e.g.  \cite{dyson}).  

This relation between the existence of scalar profiles with screening, and the sign of the higher dimensional operator is reversed when we consider homogeneous  but time dependent configurations.  Consider, for example, a scenario in which the homogeneous scalar is excited from its trivial vacuum state to one of constant energy density, ${\cal E}>0$. Then the homogeneous field equations read
\be
\frac12 \dot \varphi^2+\frac34 \epsilon \frac{\dot \varphi^4}{\Lambda^4}=\cal E \, .
\ee
Now the linearized solution $\varphi_{lin}=\sqrt{2 \cal E} t$ can only be continuously  extended to much higher energy configurations when $\epsilon=+1$.  This is important since we have screening for ${\cal E} \gg \Lambda^4 $ whence $Z \sim \sqrt{\cal E}/\Lambda^2 \gg 1$. Here the scale $\Lambda$ doubles up as both the Vainshtein scale and the vacuum strong coupling scale\footnote{This seems qualitatively different to what happens in the static spherically symmetric configurations with screening, for which the Vainshtein radius $r_V \gg 1/\Lambda$. However, even in those scenarios, the gradient energy density stored in the field at the Vainshtein radius is of order $\varphi'^2(r_V) \sim \left(\frac{M}{M_{pl}r_V^2 }\right)^2\sim \Lambda^4$.}.
If we were to now extract an environmental strong coupling scale based on fluctuations about this energetic background we would find that for high energies (${\cal E} \gg \Lambda^4$), the environmental strong coupling scale is given by $\Lambda_{\text{env}} \sim {\cal E}^{1/4} \gg \Lambda$.
In contrast, for $\epsilon=-1$, the homogeneous solution runs into a branch cut for ${\cal E}\sim \Lambda^4$ and there is no screening.

Thus, whatever the sign of $\epsilon$, we can always find some sort of Vainshtein effect, be it on static or homogeneous backgrounds.  However, analyticity considerations  suggest there is only one choice of sign ($\epsilon=+1$) that can be consistently extended into the UV beyond the cut-off $\Lambda$ by integrating out weakly coupled physics in the usual way \cite{uvir} (although see also \cite{Shore}). In this instance, a possible UV completion exists in terms of a $U(1)$ $\sigma$-model \cite{uvir}
\be \label{uv}
{\cal L}=-(\del \rho)^2-\rho^2 (\del \alpha)^2-\lambda (\rho^2-\eta^2)^2.
\ee
Here the mass of the radial mode  $m^2 =4 \lambda \eta^2$. Below this scale we can integrate out the radial mode. The low energy effective description is given by \cite{cliff}
 \be
{\cal L}=\eta^2 \left[ -(\del \alpha)^2 +\frac{(\del \alpha)^4}{m^2}+m^2{\cal O}\left(\frac{\del^6}{m^6}\right) \right].
\ee
Now canonically normalizing $\alpha=\frac{\varphi}{\sqrt{2} \eta}$, we obtain
\be 
{\cal L}=X+ \frac{X^2}{\Lambda^4}+\ldots
\ee
where $\Lambda=\sqrt{\eta m}=\frac{m}{\sqrt{2} \lambda^{1/4}}$ so that at weak coupling $m <\Lambda$. Note that the stability of the Higgs potential ($\lambda>0$) forces us to take $\epsilon=+1$ as claimed earlier. Because of this we will now focus on Vainshtein screening for homogeneous backgrounds.

The question we would like to ask is:  for the same initial data, when do classical homogeneous solutions, $\dot \alpha_{UV}(t)$, to the UV theory (\ref{uv}) deviate from the solutions, $\dot \alpha_{IR}(t)$, to the following truncated low energy effective theory:
 \be \label{ir}
{\cal L}=\eta^2 \left[ -(\del \alpha)^2 +\frac{(\del \alpha)^4}{m^2} \right].
\ee
By ``the same initial data'', we mean the same initial conditions for $\dot \alpha$ in both cases (let $\dot \alpha_{IR}(0)=\dot \alpha_{UV}(0)=\omega$), and the same total energy density, $\cal E$. For the UV theory we also need to specify initial conditions for $\rho$. Now a low energy observer ignorant of the details of the UV completion has no way of knowing the initial data, $\dot \rho(0)=v$ for the velocity of the  radial mode, so we scan over all the allowed possibilities $v^2  \in [0, v_{max}^2]$, for some $v_{max}$ to be identified shortly. For a given $v$, the initial data for $\rho(0)=\rho_0$ can be inferred from the matching of energy density. We scan over all allowed initial data by  introducing a parameter $\theta \in [0,\pi/2]$ and setting $v^2= v_{max}^2 \sin^2 \theta$.

The dynamics of the low energy effective theory is trivial: $\dot \alpha_{IR}(t)=\omega$ at all times. Furthermore, the energy density of this solution is  given by $
{\cal E}=\eta^2 \omega^2 +\frac{3}{4\lambda}\omega^4 $. The dynamics of the UV theory is marginally more complicated:
\ba \label{UVeqs}
\dot \rho^2+\rho^2 \dot \alpha_{UV}^2+\lambda(\rho^2-\eta^2)^2={\cal E}\, \nonumber \\
\rho^2 \dot \alpha_{UV}=J,
\ea
where the constant angular momentum is given in terms of the initial data as $J=\rho_0^2 \omega$. Indeed, we find that the motion is governed by an effective potential: $\dot \rho^2+V_{eff}(\rho)=\cal E$, where 
\be
V_{eff}(\rho)=\frac{w^2 \rho_0^4}{\rho^2}+\lambda(\rho^2-\eta^2).
\ee
As usual, the angular momentum generates a potential barrier at small radii.
To identify $v_{max}$, we evaluate the energy equation (\ref{UVeqs}) at $t=0$, yielding
\be \label{t=0}
v^2={\cal E}-\rho_0^2 \omega^2-\lambda(\rho_0^2-\eta^2)^2.
\ee

\begin{figure}
\begin{center}
\includegraphics[width=1\linewidth]{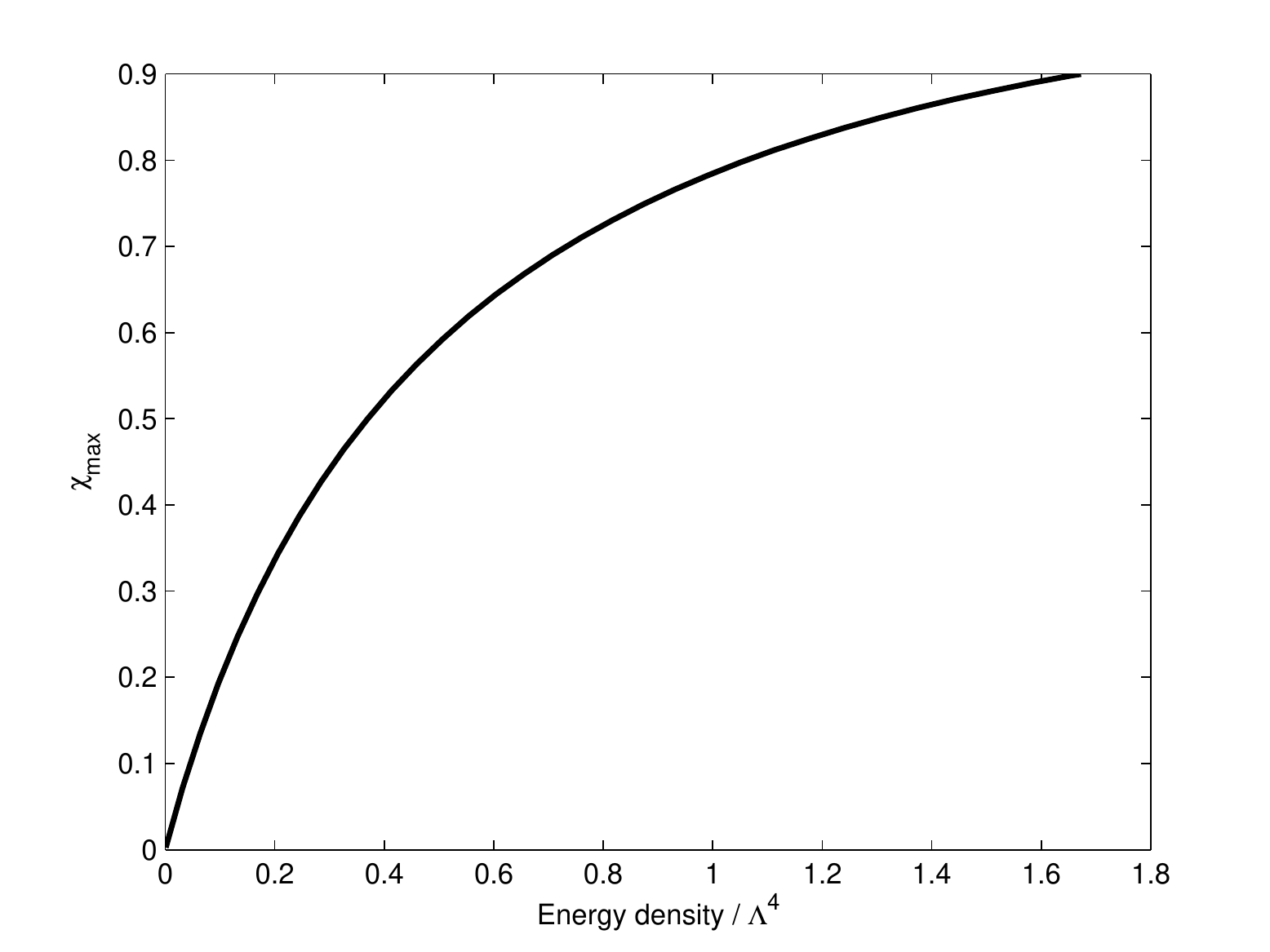}
\caption{The maximum value of $\chi$, considering a wide range of initial data for the radial mode, vs the total energy density in units of the vacuum strong coupling scale}
\label{fig:plot}
\end{center}
\end{figure}

Demanding that the RHS above be maximized as a function of $\rho_0$, we can show that $v_{max}^2=\frac{\omega^4}{\lambda}$. This represents the maximum amount of kinetic energy that can be stored in the radial mode on the initial surface, for a given angular velocity $\omega$. 
For a given $\sin^2\theta=v^2/v_{max}^2 \in [0,1]$, we can extract the initial value $\rho_0$ by solving Eq. \ref{t=0}. Because this equation is quadratic in $\rho_0^2$, there are  in general two roots. However, the smaller of these can never  be  positive (as required for real $\rho_0$) at high energies  (${\cal E} \gg \Lambda^4$, or equivalently $w^2 \gg \lambda \eta^2$) for any $\theta \in [0,\pi/2]$ so we neglect it. Instead we focus on the larger root
\be
\rho_0=\eta \sqrt{1+\frac{\omega^2}{\lambda \eta^2}\left(\cos\theta-\frac12\right)},
\ee
which always yields a real radius for some $\theta \in [0, \pi/2]$.
Now that we have set up the initial data in the UV theory we can evolve the solution for $\dot \alpha_{UV}(t)$ and compare it to our low energy solution $\dot \alpha_{IR}(t)=\omega$. For a given total energy density $\cal E$ , we propose the following measure for comparing the two theories
\be \label{chi}
\chi({\cal E})=\text{max} \left\{\left| \frac{\dot \alpha_{UV}^{2}(t)-\dot \alpha_{IR}^{2}(t)}{\dot \alpha_{UV}^{2}(t)+\dot \alpha_{IR}^{2}(t)}\right|, t\geq 0 , 0 \leq \theta\leq \pi/2\right\}.
\ee
If $\chi ({\cal E}) \gtrsim {\cal O}(0.1)$ for a given energy density ${\cal E}$, we deem the low energy solution  to be a poor approximation to the full UV solution at those energies.  In Fig \ref{fig:plot}, we plot $\chi$ against ${\cal E}/\Lambda^4$, where $\Lambda \sim \eta \lambda^{1/4}$ is the vacuum strong coupling scale.  We immediately see that   $\chi ({\cal E}) \gtrsim {\cal O}(0.1)$ whenever ${\cal E} \gtrsim \Lambda^4$.  This brings us to our main conclusion: the classical solutions to the low energy effective action (\ref{ir}) cannot be trusted in the region where Vainshtein screening is meant to take place, ${\cal E} \gg \Lambda^4$, rendering the environmental strong coupling scale dubious.

\section{Vainshtein $\&$ Higgs: $SU(2)$} \label{sec:su2}

\AP{We now turn our attention to the low energy limit of a weakly coupled non-abelian massive gauge theory. Motivated by the helicity-0 sector of massive $SU(2)$, we consider a class of Lagrangians given by}
\be \label{Y}
{\cal L}=Y+\epsilon \frac{Y^2}{\Lambda^4},
\ee
where\footnote{This comes from $Y=-\del\Phi^\dagger\del\Phi$, with the parameterization \be\nonumber\Phi=\eta\left(\begin{array}{c}\sin(\varphi/\eta)e^{i\psi_1/\eta}\\ \cos(\varphi/\eta)e^{i\psi_2/\eta}\end{array}\right)\ee.} $Y=- (\del \varphi)^2-\sin^2\left(\frac{ \varphi}{\eta}\right) (\del \psi_1^2)-\cos^2 \left(\frac{\varphi}{\eta}\right) (\del \psi_2^2)$. 
If the $\psi_i$ are not allowed to fluctuate this reduces to the scenario discussed in the previous section, so it should come as no surprise that a UV completion of the theory can  be found only when $\epsilon=+1$. This completion is, of course,  the $SU(2)$ $\sigma$-model
\begin{multline} \label{uvsu2}
{\cal L}=-(\del \rho)^2-\rho^2 \left[\del \alpha)^2+\sin^2\alpha (\del \beta_1)^2+\cos^2\alpha (\del \beta_2)^2\right] \\-\lambda (\rho^2-\eta^2)^2.
\end{multline}
Again, if we integrate out the radial mode below the mass scale $m^2=4\lambda \eta^2$, we are left with the following low energy effective theory describing the Goldstone modes 
\begin{multline} \label{irsu2}
{\cal L}=\eta^2 \left.\Bigg\{ -\left[(\del \alpha)^2+\sin^2\alpha (\del \beta_1)^2+\cos^2\alpha (\del \beta_2)^2\right] \right. \\\left. +\frac{\left[(\del \alpha)^2+\sin^2\alpha (\del \beta_1)^2+\cos^2\alpha (\del \beta_2)^2\right]^2}{m^2} \right\}.
\end{multline}
Canonically normalizing $\alpha=\varphi/\eta, ~\beta_i=\psi_i/\eta$ gives the action (\ref{Y}) with $\epsilon=+1$ and $\Lambda= \sqrt{\eta m}=\frac{m}{\sqrt{2} \lambda^{1/4}}$, as before. Assuming weak coupling in the UV theory gives $\Lambda<\eta$, and so we may assume that perturbative unitarity breaks down for the low energy theory at $\Lambda$. Furthermore, in direct analogy with the previous section, we  find Vainshtein screening (i.e. large $Z$ factors) on homogeneous backgrounds whenever the energy density of the solution ${\cal E} \gg \Lambda^4$.  

\begin{figure}
\begin{center}
 \includegraphics[width=1\linewidth]{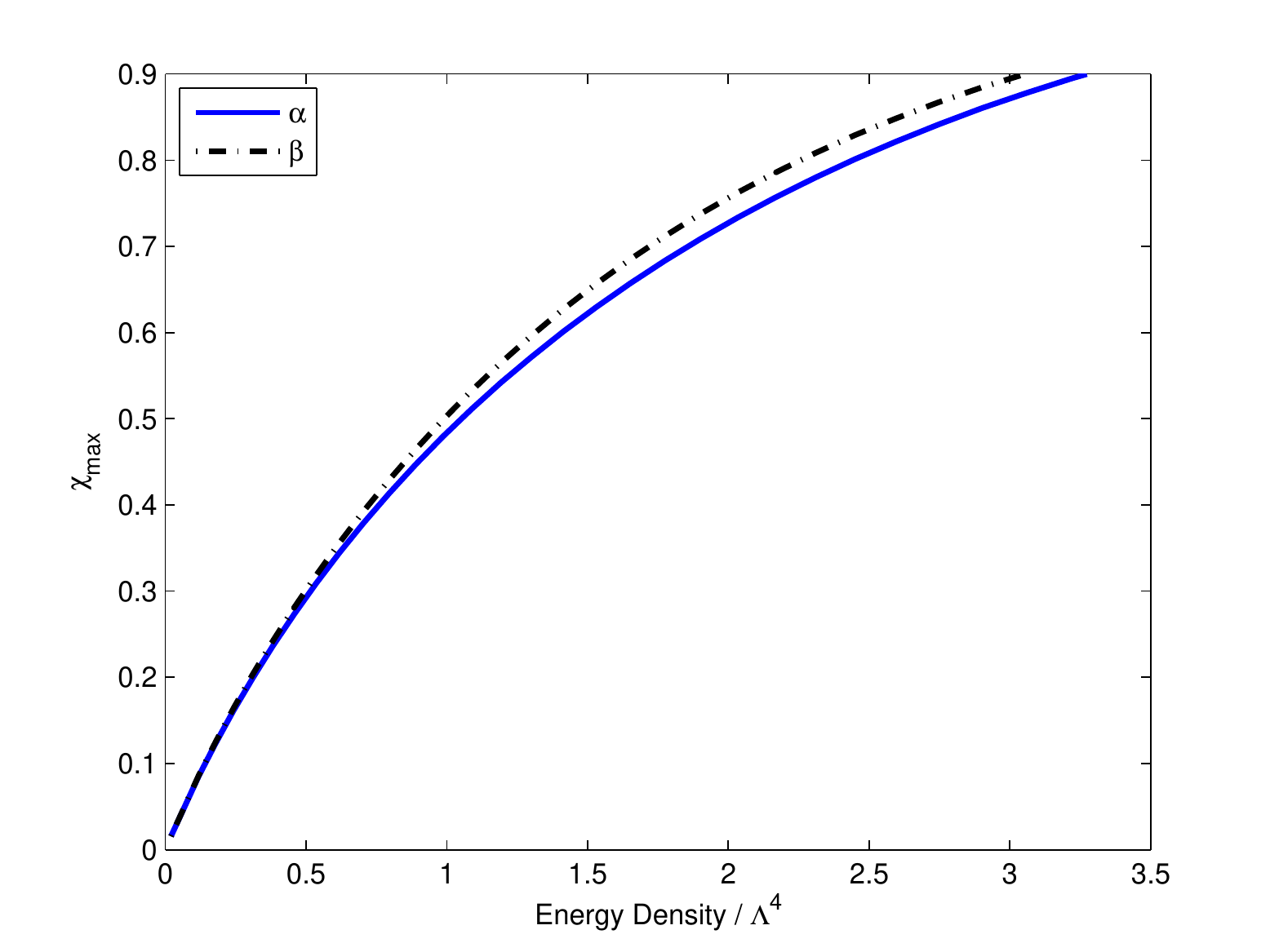}
\caption{The maximum value of $\chi$, for $\alpha$ and $\beta$, considering a wide range of initial data for the radial mode, vs the total energy density in units of the vacuum strong coupling scale}
\label{fig:plot2}
\end{center}
\end{figure}

Once again we ask whether the classical low energy solution is a good approximation to the classical solution of the UV complete theory at the scale when screening is meant to occur. To answer this we apply similar methods to  the previous section, matching initial data in the two theories, and energy density and then scanning over the allowed initial data for the radial mode. For simplicity we concentrate on the solutions for $\alpha$ and $\beta_{1}$ only, and use the same definition of $\chi(\cal E)$ for both modes as defined in (\ref{chi}). Summarizing results, Fig \ref{fig:plot2} shows that  the low energy classical solution, for both $\alpha$ and $\beta_{1}$, ceases to be a good approximation whenever ${\cal E} \gtrsim \Lambda^4$. Again, this indicates that the environmental strong coupling scale is highly questionable.   

\section{Vainshtein screening and Einstein Gauss-Bonnet gravity}  \label{sec:gb}

Let us now return to the case of Vainshtein screening in gravity. We wish to stress that in fact there is an analogue of the Vainshtein radius in General Relativity: the Schwarzschild radius. To see it, consider the perturbative description of static, spherically symmetric configurations in $D$ dimensions \cite{Duff}. Expand the metric and the Einstein-Hilbert action about Minkowski space, $g_{\mu\nu}=\eta_{\mu\nu}+h_{\mu\nu}$. Schematically,
\ba
&&\frac{M_D^{D-2}}{2}  \int d^D x \sqrt{-g} R \sim \nonumber \\
&&~~~~~~~~~~~~ \int d^D x M_D^{D-2} \left[h\del^2 h +h^2 \del^2 h +\ldots\right] \, . ~~~~~~~~
\ea
with a coupling to a source given by $\int d^D x h_{\mu\nu} T^{\mu\nu}$. The potential due to a source of mass $M$ in perturbation theory is obtained by the resummation of the  diagrams
\begin{equation}  \label{potGR}
\unitlength=1mm
\begin{fmffile}{GRpics}
\parbox{15mm}{
\begin{fmfgraph*}(15,15)
\fmfleft{pl}
\fmfright{pr}
\fmf{wiggly}{pl,pr}
\fmfblob{0.2w}{pl}
\end{fmfgraph*} } = \quad
%
%
\parbox{15mm}{
\begin{fmfgraph*}(15,15)
\fmfleft{pl}
\fmfright{pr}
\fmf{wiggly}{pl,pr}
\fmfdot{pl}
\end{fmfgraph*} } + \quad
%
%
%
%
\parbox{15mm}{
\begin{fmfgraph*}(15,15)
\fmfleft{p1,p2}
\fmfright{pr}
\fmf{wiggly}{p1,pl}
\fmf{wiggly}{p2,pl}
\fmf{wiggly}{pl,pr}
\fmfdot{p1}
\fmfdot{p2}
\end{fmfgraph*} } +  \ldots
\end{fmffile}
\end{equation}
After canonically normalizing the graviton field $h_{\mu\nu}\sim M_D^{-\frac12(D-2)} \hat h_{\mu\nu} $, we see that  each external source yields a factor of $M/M_D^{\frac12(D-2)}$, while each $3$-point vertex brings in a factor of $M_D^{-\frac12(D-2)}$. Thus the first diagram yields the usual Newtonian potential $\hat h_{lin}^c\sim\frac{ M}{M_D^{\frac12(D-2)}}r^{-(D-3)}$, while the second one yields a correction $\Delta \hat h^c \sim \frac{M^2}{M_D^{\frac32(D-2)} }r^{-2(D-3)}$. The perturbative expansion breaks down when $\Delta \hat h^c/\hat h^c_{lin} \sim 1$, or equivalently when 
\be
r \sim r_s \sim \frac{1}{M_D}\left(\frac{M}{M_D}\right)^{\frac{1}{D-3}} \, .
\ee
This is precisely the Schwarzschild radius of the source, which is indeed where the linearized perturbation theory breaks down, and one needs to consider nonlinear corrections. In standard GR, they - thanks to full $4D$ diffeomorphism invariance of the theory - add up to the full nonlinear Schwarzschild geometry. The explicit demonstration of this is nontrivial, as found by Duff \cite{Duff} and requires regulating the computations using extended sources, where the environmental effects help at sub-Schwarzschild/Vainshtein scales 
(a note of this has been made by Arnowitt, Deser an Misner in 1960 \cite{adm}!). The `strong coupling' at the `Vainshtein/Schwarzschild' radius then turns out to be merely a coordinate singularity, arising from large blueshifts experienced by infalling observers relative to static observers far away. 
In this particular example the `classical low energy theory'  (linearized GR + leading order non-linear interaction,\AP{ $\sim h^2 \del^2  h $}) is properly completed into a `fully nonlinear interacting theory' (GR)
without any new degrees of freedom ever appearing below the cutoff of the nonlinear theory (aside from the local matter modes that are needed to describe extended sources, but are already present in the theory). However, this occurs {\it because} the full diffeomorphism invariance of GR coupled to matter - a gauge symmetry - precludes any new degrees of freedom until the Planck scale. Note, that the key ingredient of this `taming' of the perturbative expansion is played by the systematic reinterpretation of the perturbative terms as corrections of the background in the boosted local 
frame, since diffeomorphism invariance ensures the universality of the divergent pieces. The expansion is regulated by taking extended sources, which are also diffeomorphism-invariant (ie, obeying stress-energy conservation \cite{adm}), and the series is then properly resummed yielding the Schwarzschild background. In general, such protection mechanisms that extend the standard $4D$ diffeomorphism invariance are not generically known in commonplace modified gravity frameworks that go beyond GR, and it remains unclear whether they exist.

\AP{However, although GR allows us to extend the classical description beyond the Schwarzschild/Vainshtein radius, it too runs into a strong coupled regime at the scale of the $D$ dimensional Planck mass, $M_D$. If we now think of GR itself as our `classical low energy theory'  then $M_D$ plays the role of the vacuum strong coupling scale, and so we now seek to ask what effect any (partial) UV complete has on the classical geometry.} In the absence of a detailed, UV complete, theory of quantum gravity, but taking string theory as the most promising candidate for it, we would expect various higher dimension irrelevant operators to correct the low energy $D$ dimensional Einstein-Hilbert action. Among them, the Gauss-Bonnet term appears to be generic \cite{GB},
\be
\frac{M_D^{D-2}}{2} \alpha' \int d^D x \sqrt{-g}  \left[ R_{\mu\nu\alpha\beta}R^{\mu\nu\alpha\beta}-4 R_{\mu\nu}R^{\mu\nu}+R^2\right],
\ee 
where the slope parameter $\alpha'$ is positive and of the order $l_s^2$, and the string length is $l_s \gtrsim 1/M_D$. In $D>4$ dimensions, black hole solutions to Einstein-Gauss-Bonnet gravity  \cite{stdeser} are given by a Schwarzschild like metric, $ds^2=-V(r) dt^2+\frac{dr^2}{V(r)}+r^2 d\Omega_{D-2}$ with a potential
\be
V(r)=1+\frac{r^2}{2\tilde \alpha'}\left[1-\sqrt{1+\frac{8 \tilde  \alpha'M}{(D-2) \Omega_{D-2}M_D^{D-2}r^{D-1}}}\right],
\ee
where $\tilde \alpha'=(D-3)(D-4)\alpha'$, and $\Omega_{D-2}$ is the volume of the unit $D-2$ sphere. If we expand the square root, then to leading order we recover the Schwarzschild solution of $D$ dimensional General Relativity 
\be
V(r) \approx 1-\frac{2M}{(D-2) \Omega_{D-2}M_D^{D-2}r^{D-3}} +\ldots 
\ee
However, this expansion is only valid if $\frac{8 \tilde  \alpha'M}{(D-2) \Omega_{D-2}M_D^{D-2}r^{D-1}}<1$, or in other words it breaks down at a scale
\be
r_{GB} \sim \frac{1}{M_D}\left(\frac{M}{M_D}\right)^{\frac{1}{D-1}} (\alpha' M_D^2)^{\frac{1}{D-1}} \gtrsim \frac{1}{M_D}\left(\frac{M}{M_D}\right)^{\frac{1}{D-1}}.
\ee
The point we want to emphasize is that  for a classical source $M \gg M_D$, the UV correction to the GR solution manifests itself at a macroscopic scale $r_{GB}$ which is within  the GR-Schwarzschild radius, $r_s$ (the analogue of the Vainshtein radius), but way beyond the cut-off $1/M_D$ (although $r_{GB}$ is inside the black hole horizon, so an external observer merely sees a small correction to the position of the horizon - a weak secondary hair). If this sort of behavior occurs in modified gravity scenarios with Vainshtein screening, one might see corrections to the scalar profile at some macroscopic scale within the Vainshtein radius but way beyond the cut-off $1/\Lambda$.

\section{Conclusions} \label{sec:conc}

Much of the work involving modifications of gravity refers to the Vainshtein mechanism as the key ingredient needed to tame the behavior of extra degrees of freedom in the realms where GR has been experimentally favored. However, the existing frameworks break down at very low scales, and it is unclear if they are meaningful microscopically. Thus it is sensible to test Vainshtein screening in models that are known to have a UV completion, and hence a healthy short distance behavior. In the examples we have considered - dealing with the Goldstone sectors of massive gauge theories - we have encountered concerning behavior of the low energy theories due to the degrees of freedom from the UV completion. These modes - even after being integrated out, but consistently so - leave their fingerprints in the low energy limits, and strongly affect the classical solutions in the region where Vainshtein screening is supposed to occur. This renders the classical solutions to the truncated EFTs that are used to exhibit Vainshtein phenomena questionable. As a corollary to this result, the environmental strong coupling, often taken as the reliable cut-off for the effective theory, is in fact not only unreliable, but it completely loses its meaning. In reality,  the cut-off to the truncated EFT should really be taken to be close to the Vainshtein radius before  any UV degrees of freedom have been excited. If we are to go beyond the Vainshtein radius, then we seem to have to worry about what those UV degrees of freedom are doing.

While this behavior seems to occur in the Goldstone sectors of massive gauge theories, and in Einstein-Gauss-Bonnet gravity, our evidence is limited to these examples.  However, it seems plausible that such examples are representatives of the behavior of massive gravity, and while we cannot be certain of it, we do think that in all likelihood the spirit of our result extends beyond the specific scenarios we have considered. In any theory with Vainshtein screening there is a breakdown of perturbative unitarity at some low scale $\Lambda$ due to some higher dimensional operators. To unitarize the theory we expect new physics to come in at some scale $m \lesssim \Lambda$ -- barring some new powerful gauge symmetry analogous to diffeomorphism invariance in GR. Such symmetries do not appear to be known at the present time. So the breakdown of the calculability will manifest in perturbaton theory 
as a tower of higher dimensional operators,  {\it in addition} to the ones included in the original theory, suppressed by the same scale $m$. Such new terms would correct both the background and the estimates of the strong coupling scale. Indeed, in the truncated EFT, Vainshtein screening kicks in when the higher dimensional operators that are retained become important. Generically the additional operators not included in the original theory will become important at a scale somewhere between the Vainshtein scale and $m<\Lambda$, in effect lowering the cutoff. Unless such effects can be reliably suppressed, the benefits of the environmental suppression of strong coupling effects may be altogether lost. More work seems to be warranted to settle this question.

\vskip.35cm

{\bf Acknowledgments}:  We would like to thank Marco Crisostomi and Guido D'Amico for useful discussions.
AP is funded by a Royal Society University Research Fellowship. DS is funded by a STFC studentship.
NK is supported in part by the DOE Grant DE-SC0009999.


\end{document}